\documentclass[conference]{IEEEtran}
\IEEEoverridecommandlockouts
\usepackage[T1]{fontenc}
\usepackage[utf8]{inputenc}
\usepackage{graphicx}
\usepackage{hyperref}
\usepackage{amsmath}
\usepackage{amsfonts}
\usepackage{amssymb}
\usepackage{cite}
\usepackage{amsmath,amssymb,amsfonts}
\usepackage{algorithmic}
\usepackage{graphicx}
\usepackage{textcomp}
\usepackage{xcolor}
\usepackage{enumitem}
\begin{document}

\title{An Adaptive System Architecture for Multimodal Intelligent Transportation Systems}
\author{\IEEEauthorblockN{Muhammad~Farooq\IEEEauthorrefmark{1}, Nima~Afraz\IEEEauthorrefmark{1}, and Fatemeh~Golpayegani\IEEEauthorrefmark{1}}\IEEEauthorblockA{\IEEEauthorrefmark{1}School of Computer Science,
University College Dublin, Ireland\\
 Email: mfarooq@ieee.org; nima.afraz@ucd.ie; fatemeh.golpayegani@ucd.ie}}
\maketitle

\begin{abstract}
Multimodal intelligent transportation systems (M-ITS) encompass a range of transportation services that utilise various modes of transport and incorporate intelligent technologies for enhanced efficiency and user experience. There are several challenges in M-ITS including data integration, Interoperability, scalability, user experience, etc. To address these challenges, such a system requires an adaptive system architecture that enables M-ITS to operate as an integrated ecosystem. In this paper, we provide an adaptive, user-centric, and layered architecture for multimodal transportation systems. The proposed architecture ensures scalability for seamless interactions of various subcomponents, that are often managed by different stakeholders. Concurrently, the data architecture is detailed, covering diverse data sources, advanced analytics, and stringent governance, providing a robust basis for intelligent decision-making. We provide two example use cases of the proposed architecture, showing how the data architecture and the system architecture can be fused and serve multimodal intelligent transport services.
\end{abstract}
\section{Introduction}
\subsection{Background of multimodal transportation system}
In the dynamic landscape of contemporary urban transportation, the emergence of multimodal ecosystems has reshaped the way individuals travel and interact with their cities. The intricate web of interconnected modes of transportation, known as multimodal ecosystems, integrates various means of transport, offering users a thorough and flexible array of choices for their trips. The background of this multimodal ecosystem lies in the growing need for sustainable and efficient transportation solutions that address the challenges brought by urbanization, population growth, and environmental concerns. Such solutions and services require an integrated and adaptive architecture that enables different systems to interact with each other and complement their services \cite{golpayegani2020intelligent}. An adaptive system architecture refers to a flexible and dynamic framework designed to accommodate changing requirements, conditions, and environments effectively. 

\begin{figure*}
 \centering
 \includegraphics[width=0.72\linewidth]{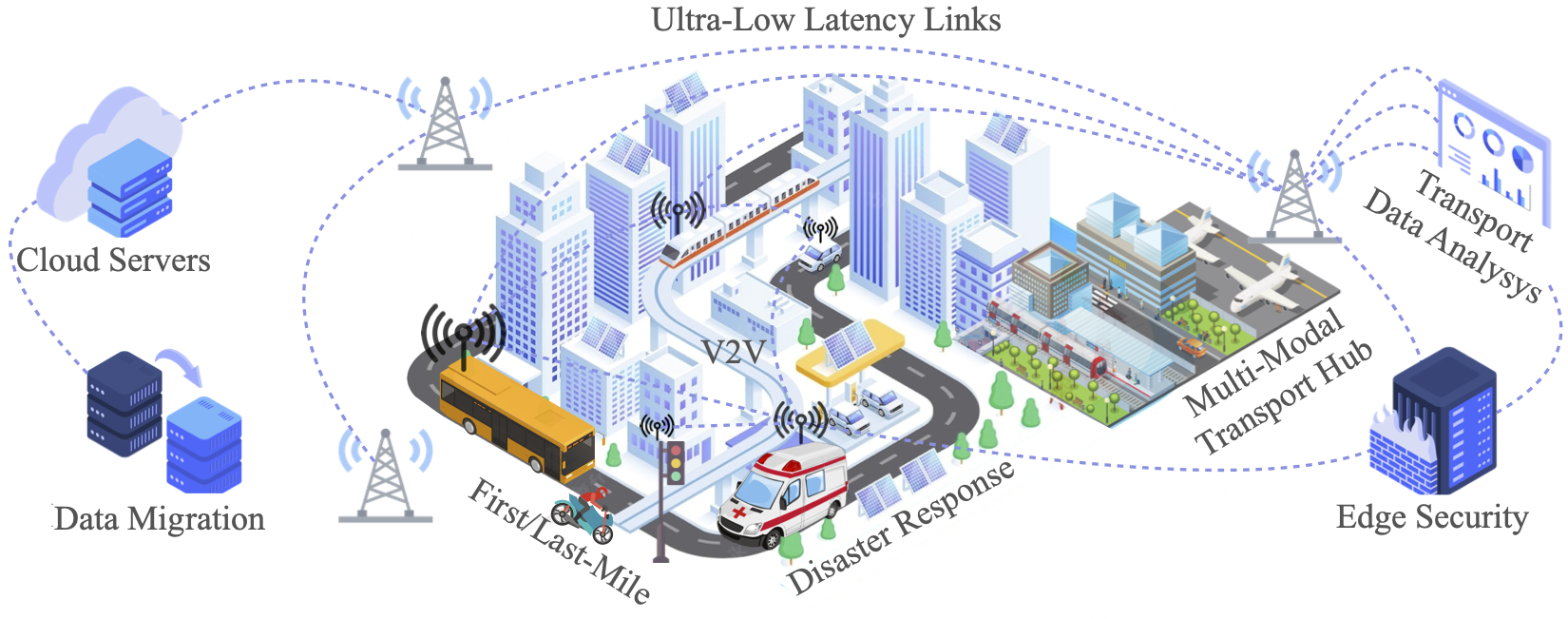}
 \caption{An example of an adaptive multimodal transportation system}
 \label{fig:MMsystem}
\end{figure*}

An example of an adaptive multimodal transportation system is shown in Fig. \ref{fig:MMsystem}. In the figure, ultra-low latency links provide communication services that can be made adaptive to wired, wireless, or vehicular ad hoc network (VANET) communication. The road users and modes of transport can also be seen in the figure. The modes of transport communicate with one another through vehicle-to-vehicle (V2V) links, and they communicate with the infrastructure using vehicle-to-infrastructure (V2I) links. The cloud and edge servers have been shown to cater to the computation services, which can be adaptable to the computational requirements. An adaptive architecture for multimodal transport systems is crucial for ensuring efficient first and last-mile connectivity, particularly in scenarios like disaster response, where seamless integration of various modes of transportation is essential for swift and effective relief efforts. 

\subsubsection{Actors and players in the system }
In the world of multimodal ecosystems, various actors and players contribute to the functionality and success of the system. 
Service providers cover a spectrum of providers of transport modes and other direct services, including standard public transportation, ride-sharing platforms, bike-sharing programs, and emerging technologies like autonomous vehicles. Users, ranging from daily commuters to occasional tourists, connect with these services. Road operators and county/city councils play a pivotal role in shaping the regulatory framework and infrastructure development to support a seamless multimodal experience. This is not an exhaustive list, but it is an indicative list of actors in this complex ecosystem. These might differ depending on the availability of services in a location, the management and ownership of these services, and the infrastructure.
\subsubsection{Benefits for actors and players}
The benefits of a well-integrated multimodal ecosystem stretch to a broad spectrum of actors, users, and stakeholders. Users gain access to a diverse range of transportation choices tailored to their needs, promoting convenience and flexibility. City councils gain from reduced congestion, lower environmental impact, and better overall urban mobility. Service providers find chances for innovation and collaboration within a thriving and interconnected transportation network. In essence, the adaptive architecture of multimodal ecosystems aims to create a win-win scenario for all involved parties.
\subsubsection{Advantages of an adaptive multimodal transportation system}

An adaptive system architecture offers a range of advantages:
\begin{itemize}
\item Seamless operation of multi-actor systems:
The complexity of multimodal ecosystems needs an adaptive architecture to ensure the seamless operation of multi-actor systems. With diverse services and providers coexisting, an adaptive framework supports the integration of these elements, minimizing disruptions and building a fluid, interconnected transportation network \cite{ter2006task}. This adaptability is crucial for maintaining the functionality of the ecosystem in the face of changing user patterns, technological advances, and evolving urban landscapes.
\item Interoperability of the systems: Interoperability sits at the heart of an effective multimodal ecosystem. An adaptive architecture allows different transportation modes and services to interact and collaborate seamlessly. This interoperability ensures that users can easily transition between modes, combining various services for a unified travel experience \cite{agbaje2022survey}. 
It breaks down silos, creating a cohesive transportation network where each component complements the others, enhancing the overall efficiency of the system.
\item Integrated Resilience in the system:
The adaptability of a design is vital to the integrated resilience of multimodal systems. In the face of unforeseen events, such as natural disasters, infrastructure problems, or sudden changes in demand, an adaptive system can quickly adjust and recover \cite{rak2021disaster}. This resilience ensures the continued availability and functionality of transportation services, adding to the reliability and sustainability of the multimodal ecosystem.
\end{itemize}
\subsubsection{A Case study: Seamless Urban Commutation} To illustrate the significance of multimodal system architecture in multimodal ecosystems, consider a bustling metropolis where commuters smoothly transition from bicycles and buses to ride-sharing services and subway systems in their daily journeys. This complex, interwoven network of diverse transportation modes exemplifies the need for an adaptive design that can harmonise the different elements of the multimodal system, providing users with a cohesive and user-friendly travel experience. In this example, the multi-modal trip includes multiple stakeholders, service providers, and platforms. For a seamless service, all of these entities must be able to interact, integrate, and offer adaptive services that can cater to the needs of users, and cope with changes in the system, technology, and new services.

\begin{figure*}[tbh]
\centering
\includegraphics[width=0.72\linewidth]{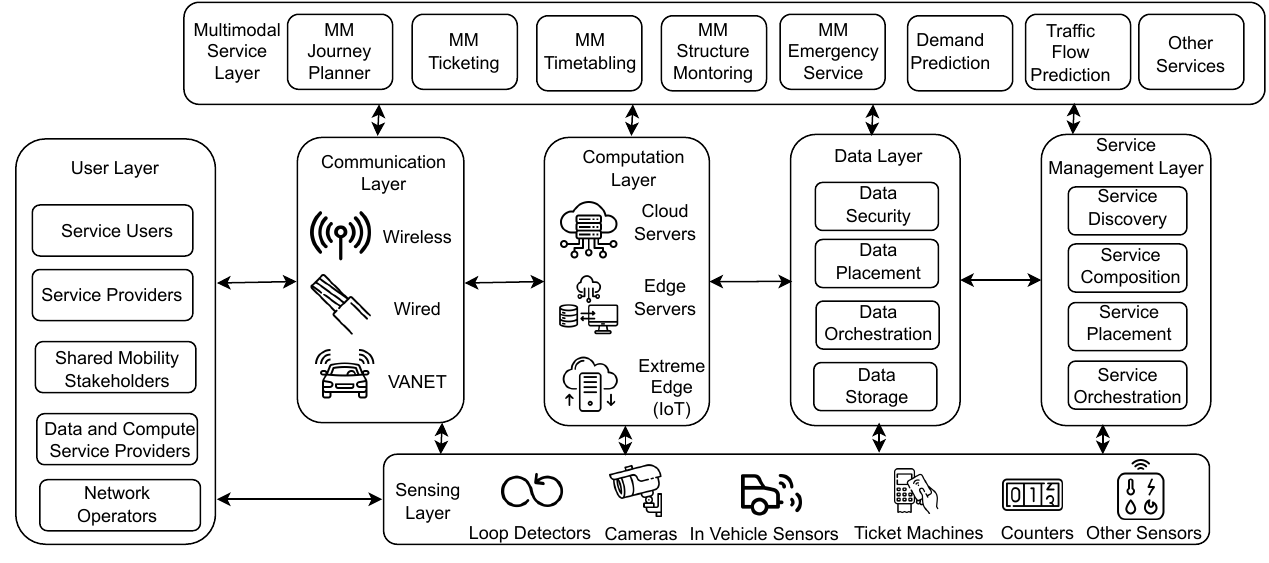}
\caption{Component level architecture for multimodal transport}
\label{Fig:comparch}
\end{figure*}

\subsection{Literature Review}
Multimodal transportation systems accommodating various transportation modes are characterized by intricate dynamics, highlighting the pivotal role of adaptive system design in managing interactions among their components. Existing research emphasizes the significance of adaptive systems in addressing the complexities of such transportation systems.

Adaptive system architecture plays a crucial role in facilitating real-time modifications to meet the evolving demands of shared multimodal transportation systems. The integration of diverse elements within a flexible framework, including vehicle management, user interfaces, connection modules, and infrastructure, is essential for enhancing operational efficiency and user experience \cite{golpayegani2020intelligent}.

Gringeri \emph{et al.} advocate for a scalable and adaptable design to support different operational requirements \cite{gringeri2010flexible}. Ystgaard \emph{et al.} highlight the relevance of human-centric design principles in developing user interfaces and system functionalities, emphasizing the role of user experience in adaptive system architectures \cite{ystgaard2023review}.

Hamilton \emph{et al.} offer insights into the adaptive management of energy resources in shared multimodal systems \cite{hamilton2013case}. Guan \emph{et al.} explore dynamic pricing models to optimize resource usage within shared multimodal platforms \cite{guan2020towards}. Addressing security concerns, Saeed \emph{et al.} provide a comprehensive literature on frameworks for integrating advanced cybersecurity protocols in shared multimodal systems \cite{saeed2023systematic}.

García \emph{et al.} analyse the state of electric Mobility as a Service (eMaaS), providing insights into ecosystem dynamics and system architectures \cite{garcia2019state}. Pham \emph{et al.} investigate the integration of edge computing technologies, and Smith and White explore modular approaches to adaptive system architecture in shared multimodal settings \cite{pham2020survey}. Fiore \emph{et al.} concentrated on data analytics for smart urban transport management \cite{fiore2019integrated}.

Staniewska \emph{et al. }discussed sustainability frameworks for adaptive system architecture \cite{staniewska2023framework}. Turon \emph{et al.} et al. studied optimization strategies of shared e-mobility systems \cite{turon2020holistic}. D\k{a}browski \emph{et al.} emphasized on adaptive frameworks for goal-oriented strategic decision-making for e-mobility systems \cite{dkabrowski2017towards}. Rodrigues \emph{et al.} investigated optimization solutions for user experience in hybrid multimodal routing \cite{rodrigues2021exploring}.

In summary, the literature reveals that adaptability, flexibility, scalability, and user-centric design are focal points in addressing the complexities and dynamic operational requirements of multi-modal transportation systems. The system architecture should be designed with these qualities in mind to ensure adaptability at runtime.
\subsection{Contributions and paper organisation}
Our contributions to this paper are as follows:
\begin{itemize}
\item We propose an adaptive layered architecture for multimodal transportation systems that consists of adaptable user, sensing, communication, communication, computation, data, service management, and multimodal service layers. These layers of the architecture interact with one another to offer effective multimodal transportation services. 
\item The proposed system architecture consists of data from multiple sources and thus, a high amount of data needs to be stored and processed. Therefore, we also propose a detailed adaptive multimodal data architecture. Data models, sources, integration, storage and organisation, analysis and evaluation, and governance form the adaptive data architecture.
\item  We provide two use case examples of the proposed multimodal architecture: a traffic monitoring system using magnetic sensors and a traffic flow prediction using real-time traffic data. For each use case, it is shown how these systems adapt to the proposed system architecture.
\end{itemize}

\begin{figure*}[tbh]
\centering
\includegraphics[width=0.72\linewidth]{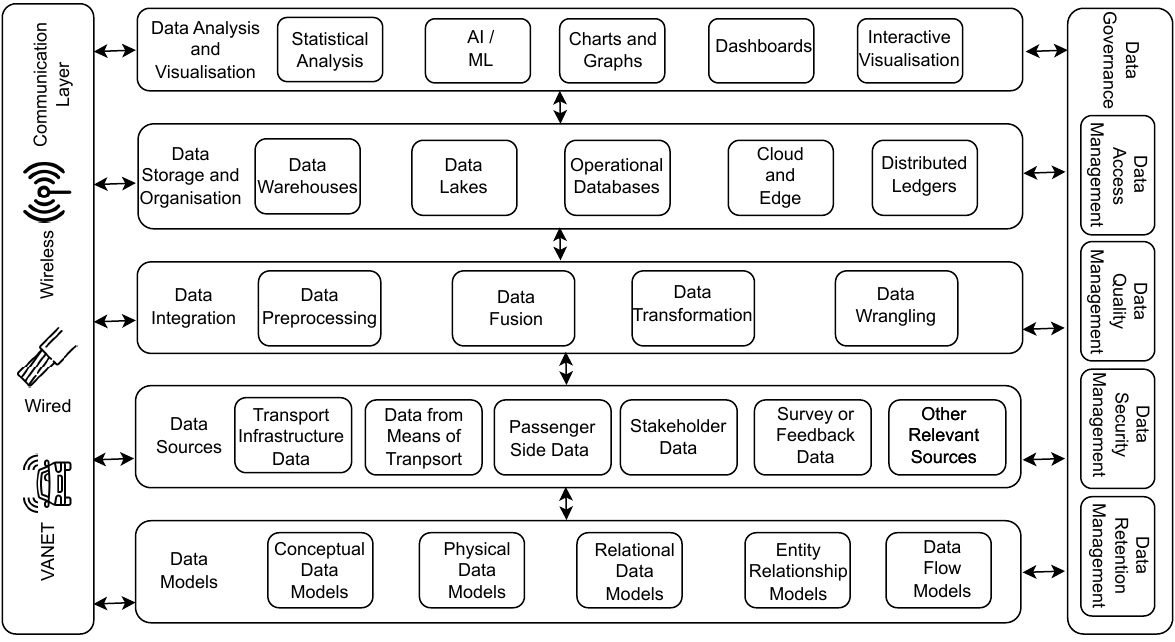}
\caption{Data level architecture for multimodal transport}
\label{Fig:dataarch}
\end{figure*}

\section{Adaptive multimodal transportation system Architecture}
We present a multimodal transportation architecture adaptable to specific demand applications. Initially, we outline the comprehensive system architecture, detailing its various layers. Subsequently, we delve into the intricacies of the data architecture, offering insights into its individual components.
\subsection{Proposed System Architecture}
The architecture diagram in Fig. \ref{Fig:comparch} illustrates a component diagram for an adaptive multimodal transportation system. The primary objective is to facilitate the use of multiple transportation means, such as buses, trains, bikes, and cars, within a single journey. The diagram showcases layered architecture encompassing all aspects of the system. The layers include the user layer, sensing layer, communication layer, computation layer, data layer, service management layer, and multimodal service layer.

These layers serve distinct purposes in the proposed multimodal transportation system. Starting with the user layer, it involves actors like service users (e.g., passengers), service providers, shared mobility stakeholders, data and compute service providers, and network operators. Each plays a vital role in utilizing and contributing to the functionality of the multimodal transportation system.

The sensing layer incorporates various sensors such as loop detectors, cameras, in-vehicle sensors, ticket machines, and other sensors. These sensors contribute to comprehensive data collection, including traffic flow, incidents, vehicle information, and environmental conditions.

Moving to the communication layer, it involves wireless and wired communication methods, ensuring real-time data transmission between sensors, devices, and the system. Additionally, the Vehicle Ad-hoc Network (VANET) facilitates direct communication between automobiles, enhancing connectivity.

The computation layer comprises cloud servers, edge servers, and extreme edge (IoT) devices. Cloud servers enable centralized and scalable data analysis, processing, and storage. Edge servers, located closer to data sources, process requests in real-time with lower latency. IoT devices contribute to distributed processing at the edge.

The data layer focuses on data security, placement, orchestration, and storage. It emphasizes safeguarding sensitive data, optimizing data placement across cloud, edge, and IoT devices for balanced performance, orchestrating data processing for scalability and flexibility, and ensuring safe and dependable storage.

The service management layer involves service discovery, composition, placement, and orchestration. These processes ensure effective communication, coordination of services, and maximization of efficiency in resource utilization.

Finally, the multimodal service layer encompasses various services offered by the system. These include the multimodal journey planner, ticketing, scheduling, structure monitoring, emergency services, demand prediction, traffic flow prediction, and additional features contributing to the seamless operation of the multimodal transportation system.

In summary, the proposed multimodal transportation system component architecture comprises these seven layers, each serving a specific purpose. The layered approach allows for adaptability and tailoring to specific system requirements. It provides a comprehensive framework to manage interactions, data flow, and services within a distributed and dynamic multimodal transportation environment.

\subsection{Proposed Data Architecture}
The proposed data architecture for the multimodal transportation system is depicted in Fig. \ref{Fig:dataarch}. It adopts a layered architecture consisting of data models, data sources, data integration, data storage and organization, data analysis and visualization, and data governance layers. These layers collectively contribute to the effective handling and processing of data within the multimodal transportation system.

The data architecture encompasses six layers, each serving a specific purpose in the multimodal transportation system. Starting with the Data Models layer, it includes conceptual data models, physical data models, relational data models, entity relationship models, and data flow models. These models provide high-level representations and detailed descriptions of data structures and relationships, aiding in understanding and implementation.

Moving to the Data Sources layer, it involves various data inputs crucial for the functioning of the multimodal transportation system. These sources include transport infrastructure data, real-time information from means of transport, passenger-side data, stakeholder data, survey or comments statistics, and other relevant sources. These diverse data inputs contribute to a comprehensive understanding of the transportation ecosystem.

The Data Integration layer focuses on processing and preparing the collected data. It includes data preprocessing, data fusion, data transformation, and data wrangling. These processes ensure that the data is cleaned, merged, and transformed into a usable format for further analysis.

The Data Storage and Organization layer involves the utilization of different storage methods. This includes data warehouses for storing historical traffic trends, data lakes for handling large volumes of raw data, operational databases for managing live data, and cloud and edge storage for persistency, scalability, and efficient data access.

Moving forward, the Data Analysis and Visualization layer encompasses statistical analysis, machine learning, charts and graphs, dashboards, and interactive visualizations. These components aid in interpreting historical traffic patterns, developing predictive models, and providing real-time displays of traffic situations and system performance.

Lastly, the Data Governance layer ensures the effective management and security of data within the multimodal transportation system. It includes data access management, data quality management, data security management, and data retention management. These measures collectively safeguard the integrity, security, and accessibility of the data.

In summary, the proposed multimodal transportation system architecture combines a sophisticated component architecture with a robust data architecture to model shared multimodal and multimodal transportation systems. The component architecture, characterized by its user-centric design and layered structure, facilitates seamless interactions and ensures modularity, scalability, and an improved user experience. Complementing this, the data architecture uses a comprehensive approach, encompassing diverse data sources, advanced analytics, and stringent governance. From conceptual data models to real-time data integration, the multimodal transportation system's data architecture stands as a resilient foundation, providing secure, high-quality, and adaptive control of transportation systems. 
Together, these architectures position the multimodal transportation system at the forefront of intelligent, data-driven solutions for the evolving landscape of shared transportation.
\section{Aspects of the proposed adaptive multimodal architecture}
In this section, we discuss the important aspects of the proposed adaptive architecture for multimodal transportation systems.
\subsection{Data}
In the realm of data considerations for the multimodal transportation system architecture, three important aspects merit careful attention:
\subsubsection{Data Sharing Protocols}
Effective data sharing protocols form the basis of the multimodal transportation system's adaptive architecture. Establishing standardised protocols ensures seamless interoperability among different components. Protocols such as RESTful APIs, GraphQL, and MQTT enable efficient and secure data exchange, promoting a cohesive and integrated transportation ecosystem.
\subsubsection{Data Storage}
Efficient and scalable data storage is necessary for handling the vast datasets inherent in shared multimodal and multimodal transportation systems. Employing distributed databases, both on-premises and in the cloud, allows multimodal transportation systems to handle dynamic datasets effectively, ensuring data availability, reliability, and scalability.
\subsubsection{Data Security}
Security methods within the data architecture of multimodal transportation systems are multifaceted. Implementing end-to-end encryption, access controls, and regular security audits safeguards private transportation data. Adherence to industry standards and best practices is crucial to ensure the integrity, confidentiality, and availability of data.
\subsection{Communication}
The communication aspects of the multimodal transportation system architecture spans various dimensions, addressing interactions between components, external APIs/internal systems, and end-users.
\subsubsection{Between Components of the Architecture (Shared Platform)}
Seamless communication between components is pivotal for the shared platform's operation. Utilizing messaging protocols like MQTT or event-driven architectures enhances real-time information sharing, fostering dynamic adaptability within the system.
\subsubsection{Communication Between Shared Platform and External APIs/Internal Systems}
Integration with external APIs and internal systems is eased through standardised communication protocols such as HTTP/HTTPS and message queues. This ensures compatibility, allowing multimodal transportation systems to connect with diverse services and platforms.
\subsubsection{Communication with End Users}
End-user communication includes user-friendly interfaces and real-time updates. Utilizing responsive web design, push notifications, and efficient backend communication improves the overall user experience, promoting engagement and satisfaction.
\subsection{Computation}
Computation within the multimodal transportation system architecture includes cloud computing, edge computation, and mist computing, forming a continuum of computational capabilities.
\subsubsection{Cloud Computing}
Leveraging cloud computing resources offers scalability, flexibility, and cost-efficiency. Cloud platforms such as AWS, Azure, or Google Cloud allow multimodal transportation systems to process extensive datasets and deploy complex algorithms.
\subsubsection{Edge Computation}
Edge computation brings computational resources closer to the data source, lowering latency and enhancing real-time decision-making. Edge devices, such as IoT devices on vehicles, add to processing localised data, optimizing system responsiveness.
\subsubsection{Computation at the Mist}
Mist computing includes decentralised computation at the network's edge, enhancing the scalability and reliability of data processing. Implementing mist computing in multimodal transportation systems adds to efficient data handling and decision-making in distributed environments.
\subsubsection{Cloud to Edge Continuum}
The cloud-to-edge continuum in multimodal transportation systems suggests a hybrid approach, strategically distributing computational tasks between the cloud and edge devices. This ensures optimal resource usage, responsiveness, and adaptability to changing computational demands.
\subsection{Security}
Ensuring security across different layers is important for the robustness of the multimodal transportation system architecture.
\subsubsection{Different Layers}
Security measures cover multiple layers, encompassing network security, application security, and data security. Encryption, firewalls, and intrusion detection systems work synergistically to safeguard the entire architecture.
\subsubsection{Authentication and Authorization}
Authentication and authorization methods play a pivotal role in securing access to multimodal transportation system components. Implementing multifactor authentication, role-based access controls, and secure identity management improves the overall security posture, ensuring that only authorised entities interact with sensitive systems and data.
\subsubsection{Incident Response and Recovery}
Preparedness for security incidents is important. Developing robust incident reaction and recovery plans ensures swift and effective responses to possible threats. Continuous monitoring, real-time threat intelligence, and proactive measures contribute to minimizing the effect of security incidents on the multimodal transportation system.
\subsection{Integration}
Integration considerations within the multimodal transportation system architecture cover the creation of new services and their linkage, emphasizing standardised approaches.
\subsubsection{How New Services Will be Developed and Linked}
Integrating new services within the multimodal transportation system architecture includes adopting container-based software development, standardised interfaces, data sharing protocols, and computation. The utilization of containerization streamlines service deployment, while standardised interfaces and data sharing protocols promote interoperability. multimodal transportation system's commitment to standardised computation promises consistency and compatibility in service creation and integration.
\subsubsection{Microservices Architecture}
Embracing a microservices design further enhances the flexibility and scalability of service development and integration. Decomposing the system into modular and separately deployable microservices allows for agile development, updates, and upkeep. This method aligns with multimodal transportation systems' commitment to adaptability and ease of integration.
\subsection{Standardisation}
In the setting of the multimodal transportation system, the application of Intelligent Transportation Systems (ITS) standards plays a pivotal role in ensuring interoperability, efficiency, and seamless integration within the multimodal transport ecosystem.
\subsubsection{Data Sharing}
Adherence to established ITS standards for data sharing, such as DATEX II~\cite{datex2} for traffic management information and SIRI (Service Interface for Real-time Information)~\cite{siri} for real-time public transportation data, supports a unified and standardized approach. This ensures that disparate systems can share information seamlessly, promoting comprehensive data sharing across the transportation network.
\subsubsection{Service Development}
Standardized interfaces and protocols are important for the development of new services within the multimodal transportation system architecture. Leveraging ITS standards like TMDD (Traffic Management Data Dictionary)~\cite{tmdd} for traffic data and CEN/TS 17123~\cite{cents17123} for multimodal travel information provides consistency in service development. This standardization not only promotes collaboration among diverse stakeholders but also expedites the creation and integration of innovative transportation services.

The integration of ITS standards not only aligns with industry best practices but also future-proofs the multimodal transportation system, ensuring compatibility with evolving technologies and creating a cohesive and interoperable multimodal transport environment.

\begin{figure*}[tbh]
\centering\includegraphics[width=0.72\linewidth]{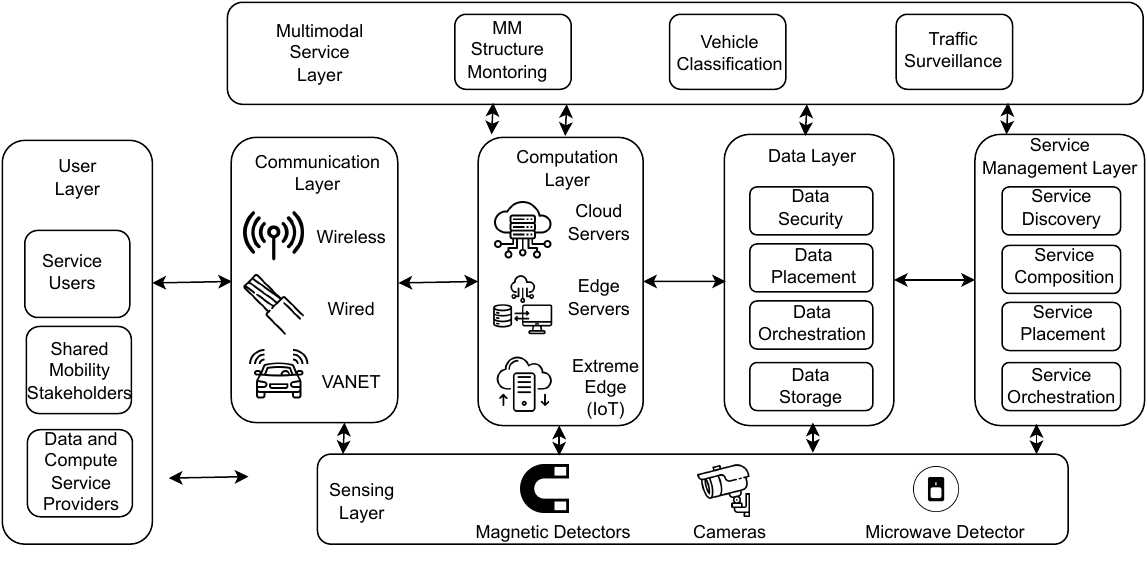} \caption{Architecture for traffic monitoring system using magnetic sensor use case}
\label{Fig:ramin}
\end{figure*}

\section{Metrics for Evaluation}\label{sec:KPIs}
Evaluating the effectiveness of an adaptive architecture within a multimodal transportation system is crucial for ensuring its efficiency, scalability, and seamless operation. Various metrics offer valuable insights into different aspects of the system, guaranteeing that it meets the requirements of users and stakeholders. The evaluation metrics for an adaptive architecture are tailored to the specific application context. Below are some key metrics for evaluating the effectiveness of a multimodal transportation system architecture:

\subsection{Performance}
Performance metrics assess the responsiveness, efficiency, and overall effectiveness of the system in delivering services. Key performance indicators (KPIs) include:

\subsubsection{Latency}
Latency, measured as the time taken for data processing, communication, and decision-making, is critical for ensuring real-time responsiveness. In a multimodal transportation system, low latency is essential for providing timely information to users, such as real-time transit updates and traffic conditions.

\subsubsection{Throughput}
Throughput evaluates the system's capacity to handle a large volume of transactions and data, ensuring optimal efficiency during peak usage periods. For example, a ride-sharing platform must maintain high throughput to accommodate a surge in ride requests during rush hours.

\subsubsection{Scalability}
Scalability measures the system's ability to handle increasing loads and accommodate growth in users, data, and services. A scalable multimodal transportation system should seamlessly expand its capacity to support additional commuters, transportation modes, and services as demand grows.

\subsubsection{Response Time}
Response time refers to the time taken for the system to respond to user inputs, such as route planning requests or ticket purchases. A low response time enhances the user experience by providing prompt feedback and ensuring a seamless interaction with the system.

\subsubsection{Availability}
Availability measures the system's uptime and accessibility to users, ensuring uninterrupted access to transportation services and information. High availability is crucial for maintaining user trust and satisfaction, especially during peak usage periods.

\subsection{Security}
Security metrics evaluate the robustness of the system in safeguarding sensitive information and preventing unauthorized access. Key security measures include:

\subsubsection{Incident Response Time}
Incident response time measures the system's ability to identify and respond to security incidents promptly. A quick incident response minimizes the impact of security breaches and ensures the integrity and confidentiality of transportation data.

\subsubsection{Authentication and Authorization Success Rates}
Authentication and authorization success rates assess the effectiveness of security measures in verifying user identities and controlling access to sensitive transportation data and services. High success rates indicate a robust authentication and authorization process that mitigates the risk of unauthorized access.

\subsubsection{Encryption Effectiveness}
Encryption effectiveness evaluates the strength of encryption methods used to protect data during transmission and storage. Strong encryption ensures the confidentiality and integrity of sensitive transportation information, such as passenger details and payment transactions.

\subsection{Scalability and Flexibility}
Scalability and flexibility metrics assess the system's ability to adapt to changing requirements and accommodate growth. Key factors include:

\subsubsection{System Adaptability}
System adaptability measures how well the architecture can adjust to changes in user needs, technological advancements, and emerging transportation services. An adaptable architecture can easily incorporate updates and enhancements to meet evolving demands.

\subsubsection{Ease of Integration}
Ease of integration evaluates the system's ability to seamlessly incorporate new services and components, promoting interoperability and expanding its capabilities. For example, integrating third-party APIs for weather forecasts or traffic updates enhances the system's functionality and provides users with comprehensive travel information.

\begin{figure*}[tbh]
\centering\includegraphics[width=0.72\linewidth]{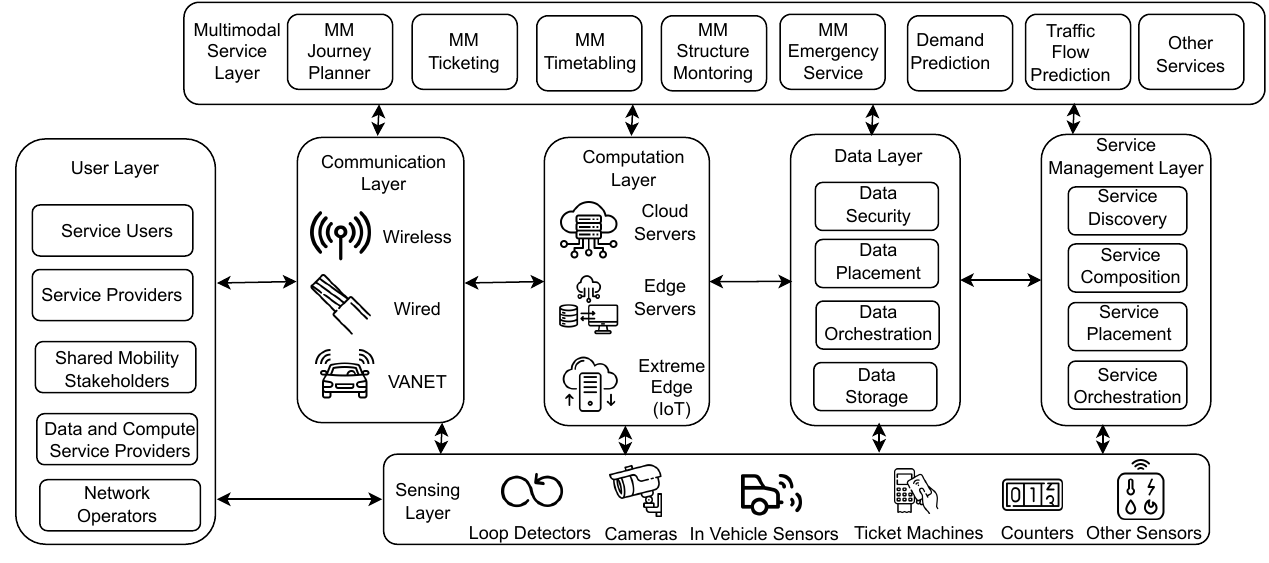} \caption{Architecture for the traffic flow prediction use case}
\label{Fig:maha}
\end{figure*}

\section{Architecture instantiation and use cases}
This section details how the proposed architecture can be instantiated using two use case examples. 
\subsection{Traffic Monitoring System using Magnetic Sensor}
The proposed architecture can be used for the development of a traffic monitoring system using a magnetic sensor. In particular, a framework can be built for on-road traffic surveillance through small and easy-to-install magnetic sensors. The magnetic sensor system is wireless-connected, cost-effective, and environmental-friendly which consists of the accelerometer and the magnetometer and is situated on the pavement. The design derived from the proposed architecture is shown in Fig. \ref{Fig:ramin}. It makes use of multimodal structure monitoring, vehicle classification, and traffic surveillance services. The sensors used for this work are magnetic detectors, cameras, and microwave detectors. The user layer for this work consists of service users, share mobility stakeholders, and data and compute service providers as shown in the figure.

Alongside the evaluation metrics proposed in Section \ref{sec:KPIs}, there are many possible metrics for the evaluation of the architecture shown in Fig. \ref{Fig:ramin} e.g., precision in detection of traffic events, latency time in the detection of events, scalability to large traffic volume, etc. One such performance metric is the correct vehicle category detection. In this case, we can set a KPI such that the architecture is validated if the correct vehicle category is detected for more than 90\% of the experiments.
\subsection{Traffic Flow Prediction}
The proposed multimodal architecture can be instantiated for the traffic flow prediction. This use case may involve analysing the average speed of vehicles on different routes at different times, and how the behaviour of the drivers affects the congestion patterns. As seen from Fig. \ref{Fig:maha}, the traffic flow prediction use case adapts to the multimodal transportation system architecture. For this purpose, multimodal journey planner, demand prediction, and traffic flow prediction services were used from the service layer of the architecture. Loop detectors and counters were the main sensors in this work. Service users or passengers make the user layer of the architecture.

There are some specific performance metrics possible for this use case too, e.g., the proportion of correctly predicted traffic flow values compared to the total number of predictions, reliability of the system for making correct traffic flow predictions, stability of prediction under various traffic conditions, etc. An example KPI for this use case is the average traffic speed passing through a traffic link normalised to the maximum speed, being more than 80\%, which represents traffic congestion in layman's terms.
\section{Conclusion}
In this paper, we have provided an adaptive architecture for multimodal transportation systems. Specifically, We have provided a carefully designed component architecture that ensures user-centricity, scalability, and modularity, fostering seamless interactions between stakeholders. Simultaneously, we have also put forward a robust data architecture, ranging from conceptual models to real-time integration, which lays the groundwork for intelligent decision-making and system optimization. We have provided two practical use case examples that make use of the proposed multimodal architecture showing how the traffic monitoring system and the traffic flow prediction use cases adapt to the proposed multimodal architecture.

\section*{Acknowledgment}

Funding for this research is provided by the European
Union's Horizon Europe research and innovation program through the
Marie Sklodowska-Curie SE grant under the agreement RE-ROUTE No 101086343.

\bibliographystyle{ieeetr}
\bibliography{references}
\end{document}